\documentclass{article}
\usepackage{amsmath,amssymb,amsthm}
\usepackage{graphicx}
\numberwithin{equation}{section}

\theoremstyle{definition}

\theoremstyle{remark}

\newcommand{\beq}{\begin{eqnarray}}
\newcommand{\eeq}{\end{eqnarray}}
\newcommand{\beqnn}{\begin{eqnarray*}}
\newcommand{\eeqnn}{\end{eqnarray*}}
\newcommand{\rd}{\partial}

\newcommand{\CC}{\mathbf{C}}
\newcommand{\PP}{\mathbf{P}}
\newcommand{\RR}{\mathbf{R}}
\newcommand{\ZZ}{\mathbf{Z}}

\begin{document}

\title{Integrable structure of melting crystal model\\
with external potentials}

\author{Toshio Nakatsu$^1$%
~and Kanehisa Takasaki$^2$%
\thanks{E-mail: takasaki@math.h.kyoto-u.ac.jp}\\
\\
$^1$Faculty of Engineering, Mathematics and Physics,Setsunan University\\
Ikedanakamachi, Neyagawa, Osaka 572-8508, Japan\\
$^2$Graduate School of Human and Environmental Studies, Kyoto University\\
Yoshida, Sakyo, Kyoto 606-8501, Japan}

\date{}

\maketitle

\begin{abstract}
This is a review of the authors' recent results 
on an integrable structure of the melting crystal model 
with external potentials. The partition function of this model 
is a sum over all plane partitions (3D Young diagrams). 
By the method of transfer matrices, this sum turns into 
a sum over ordinary partitions (Young diagrams), which may be 
thought of as a model of q -deformed random partitions. 
This model can be further translated to the language of 
a complex fermion system. A fermionic realization of 
the quantum torus Lie algebra is shown to underlie therein. 
With the aid of hidden symmetry of this Lie algebra, 
the partition function of the melting crystal model 
turns out to coincide, up to a simple factor, 
with a tau function of the 1D Toda hierarchy. 
Some related issues on 4D and 5D supersymmetric 
Yang-Mills theories, topological strings and 
the 2D Toda hierarchy are briefly discussed.
\end{abstract}

\newpage

\section{Introduction}

\begin{figure}
\begin{center}
\includegraphics[scale=0.4]{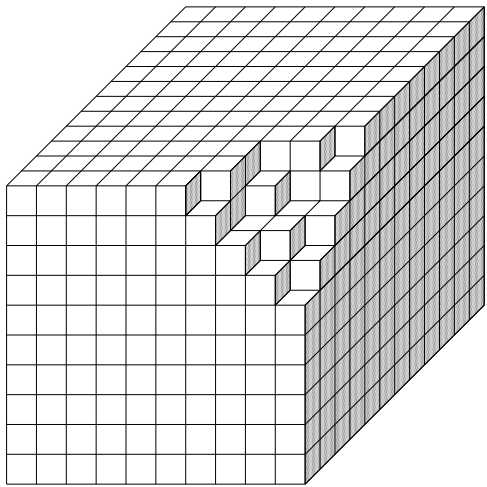}
\end{center}
\caption{Melting crystal corner}
\label{fig:crystal}
\end{figure}
\begin{figure}
\begin{center}
\includegraphics[scale=0.6]{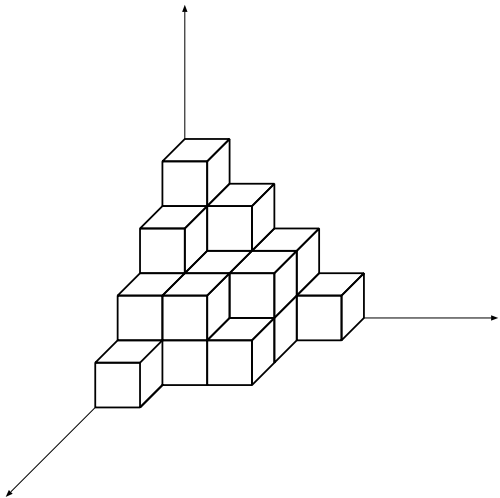}
\end{center}
\caption{3D Young diagram as complement of crystal corner}
\label{fig:3DYoung}
\end{figure}

The melting crystal model is a model of statistical 
mechanics that describes a melting corner of 
a semi-infinite crystal (Figure\ref{fig:crystal}).  
The crystal is made of unit cubes, which are 
initially placed at regular positions and 
fills the positive octant $x,y,z \ge 0$ of 
the three dimensional Euclidean space.  
As the crystal melts, a finite number of cubes 
are removed from the corner.  The present model 
excludes such crystals that have ``overhangs'' 
viewed from the $(1,1,1)$ direction.  
In other words, the complement of the crystal 
in the positive octant is assumed to be a 3D analogue 
of Young diagrams (Figure\ref{fig:3DYoung}).  
Since 3D Young diagrams are represented by 
``plane partitions'', the melting crystal model 
is also referred to as a model of 
``random plane partitions''.   

Though combinatorics of plane partitions has 
a rather long history \cite{Bressoud-book}, 
Okounkov and Reshetikhin \cite{Okounkov-Reshetikhin03} 
proposed an entirely new approach in the course 
of their study on a kind of stochastic process 
of random partitions (the Schur process).  
Their approach was based on ``diagonal slices'' 
of 3D Young diagrams and ``transfer matrices'' 
between those slices.   As a byproduct, 
they could re-derive a classical result of MacMahon 
\cite{Bressoud-book} on the generating function 
of the numbers of plane partitions.  
Actually, this generating function is nothing but 
the partition function of the aforementioned 
melting crystal model.  The method of Okounkov 
and Reshetikhin was soon generalized \cite{ORV03} 
to deal with the topological vertex \cite{Iqbal02,AKMV03} 
of $A$-model topological strings 
on toric Calabi-Yau threefolds.  

The melting crystal model is also closely related 
to supersymmetric gauge theories.   Namely, 
with slightest modification, the partition function  
can be interpreted as the instanton sum of 
5D $\mathcal{N}=1$ supersymmetric (SUSY) $U(1)$ 
Yang-Mills theory on partially compactified space-time 
$\RR^4\times S^1$ \cite{MNTT04a}.  This instanton sum 
is a 5D analogue of Nekrasov's instanton sum  
for 4D $\mathcal{N}=2$ SUSY gauge theories 
\cite{Nekrasov02,Nakajima-Yoshioka05}. 
The 4D instanton sum is a statistical sum 
over ordinary partitions (or ``colored'' 
partitions in the case of $SU(N)$ theory), 
hence a model of random partitions.  
Nekrasov and Okounkov \cite{Nekrasov-Okounkov03} 
used such models of random partitions to re-derive 
the Seiberg-Witten solutions \cite{Seiberg-Witten94} 
of 4D $\mathcal{N}=2$ SUSY gauge theories.  
Actually, by the aforementioned method of 
transfer matrices, the statistical sum 
over plane partitions can be reorganized 
to a sum over partitions.  This is a kind of 
$q$-deformations of 4D instanton sums.  
A 5D analogue of the Seiberg-Witten solution 
can be derived from this $q$-deformed instanton sum 
\cite{Nekrasov-Okounkov03,MNTT04b}. 

In this paper, we review our recent results 
\cite{Nakatsu-Takasaki07} on an integrable structure 
of the melting crystal model (and the 5D $U(1)$ 
instanton sum) with external potentials.  
The partition function $Z_p(t)$ of this model 
is a function of the coupling constants 
$t = (t_1,t_2,\ldots)$ of the external potentials.  
A main conclusion of these results is that 
$Z_p(t)$ is, up to a simple factor, a tau function 
of the 1D Toda hierarchy, in other words, 
a tau function $\tau_p(t,\bar{t})$ of 
the 2D Toda hierarchy \cite{Ueno-Takasaki84} 
that depends only on the difference $t - \bar{t}$ 
of the two sets $t,\bar{t}$ of time variables.  
To derive this conclusion, we first rewrite $Z_p(t)$ 
in terms of a complex fermion system.  
In the case of 4D instanton sum, such a fermionic 
representation was proposed by Nekrasov et al. 
\cite{Losev-etal03,Nekrasov-Okounkov03}.  
In the present case, we can use the aforementioned 
transfer matrices \cite{Okounkov-Reshetikhin03} 
to construct a fermionic representation.  
This fermionic representation, however, does not 
take the form of a standard fermionic representation 
of the (1D or 2D) Toda hierarchy \cite{Jimbo-Miwa83,Takebe91}.  
To resolve this problem, we derive a set of 
algebraic relations (referred to as ``shift symmetry'') 
satisfied by the transfer matrices and 
a set of fermion bilinear forms.  (Actually, 
these fermion bilinear forms turn out to give 
a realization of ``quantum torus Lie algebra''.) 
These algebraic relations enable us to rewrite 
the fermionic representation of $Z_p(t)$ 
to the standard form of Toda tau functions. 

In the 4D case, a similar partition function 
with external potentials has been studied by 
Marshakov and Nekrasov 
\cite{Marshakov-Nekrasov06,Marshakov07}. 
According to their results, the 1D Toda hierarchy 
is also a relevant integrable structure therein. 
Unfortunately, our method developed for the 5D case 
relies heavily on the structure of quantum torus 
Lie algebra, which ceases to exist in the 4D setup. 
We shall return to this issue, along with 
some other issue, in the end of this paper.  

This paper is organized as follows.   
Section 2 is a brief review of the melting crystal model 
and its mathematical background.  Section 3 presents 
the fermionic formula of the partition function. 
The method of transfer matrices is reviewed in detail.  
Section 4 deals with the quantum torus Lie algebra 
and its shift symmetries.  In Section 5, we use 
this symmetry to rewrite the fermionic representation 
of the partition function to the standard form 
as a Toda tau function.  Section 6 is devoted 
to concluding remarks.

\section{Melting crystal model}

\subsection{Young diagrams and partitions}

Let us recall \cite{Macdonald-book} 
that an ordinary 2D Young diagram is represented 
by an integer partition, namely, a sequence 
\beqnn
 \lambda = (\lambda_1,\lambda_2,\ldots), \quad 
  \lambda_1 \ge \lambda_2 \ge \cdots, \quad 
\eeqnn
of nonincreasing integers $\lambda_i \in \ZZ_{\ge 0}$ 
with only a finite number of $\lambda_i$'s being 
nonzero.  $\lambda_i$ is the length of $i$-th row 
of the Young diagram viewed as a collection 
of unit squares.  We shall always identify 
such a partition $\lambda$ with a Young diagram.  
The total area of the diagram is given by the degree 
\beqnn
\displaystyle |\lambda| = \sum_i \lambda_i 
\eeqnn
of the partition.  

It was shown by Euler that the generating function 
of the number $p(N)$ of partitions $\lambda$ 
of degree $N$ has an infinite product formula: 
\beq
  \sum_{N=0}^\infty p(N)q^N 
  = \prod_{n=1}^n (1 - q^n)^{-1}, 
\eeq
where $q$ is assumed to be in the range $0 < q < 1$.  
One can interpret this generating function 
as the partition function of a model of 
statistical mechanics, 
\beqnn
  Z_{\mathrm{2D}}  
  = \sum_{N=0}^\infty p(N)q^N 
  = \sum_\lambda q^{|\lambda|}, 
\eeqnn
in which each partition $\lambda$ is assigned 
an energy proportional to $|\lambda|$, and 
$q$ is related to the temperature $T$ 
as $q = e^{-\mathrm{const.}/T}$

\subsection{3D Young diagrams and plane partitions}

A 3D Young diagram can be represented by 
a ``plane partition'', namely, a 2D array 
\beqnn
\pi = (\pi_{ij})_{i,j=1}^\infty
= \left(\begin{array}{ccc}
  \pi_{11} & \pi_{12} & \cdots \\
  \pi_{21} & \pi_{22} & \cdots \\
  \vdots   & \vdots   & \ddots 
  \end{array}\right)
\eeqnn
of nonnegative integers 
$\pi_{ij} \in \ZZ_{\ge 0}$ such that 
\beqnn
  \pi_{ij} \ge \pi_{i,j+1}, \quad 
  \pi_{ij} \ge \pi_{i+1,j}. 
\eeqnn
$\pi_{ij}$ is the height of the stack of cubes 
placed at the $(i,j)$-th position of the plane.  
We shall identify such a plane partition with 
the corresponding 3D Young diagram. 
The total volume of the 3D Young diagram 
is given by 
\beqnn
   |\pi| = \sum_{i,j=1}^\infty \pi_{ij}. 
\eeqnn

As an analogue of $p(N)$, one can consider 
the number $\mathrm{pp}(N)$ of plane partitions $\pi$ 
with $|\pi| = N$.  The generating function of 
these numbers was studied by MacMahon \cite{Bressoud-book} 
and shown to be given, again, by an infinite product: 
\beq
  \sum_{N=0}^\infty \mathrm{pp}(N)q^N 
  = \prod_{n=1}^\infty (1 - q^n)^{-n}. 
\eeq
The right hand side is now called the MacMahon function.  
In statistical mechanics, this generating function 
becomes the partition function 
\beqnn
  Z_{\mathrm{3D}} 
  = \sum_{N=0}^\infty \mathrm{pp}(N)q^N 
  = \sum_{\pi}q^{|\pi|} 
\eeqnn
of a canonical ensemble of plane partitions, 
in which each plane partition $\pi$ has 
an energy proportional to the volume $|\pi|$.

We shall deform this simplest model 
by external potentials.  To this end, 
we have to introduce the notion of ``diagonal slices'' 
of a plane partition.

\subsection{Diagonal slices of 3D Young diagrams}

\begin{figure}
\begin{center}
\includegraphics[scale=0.7]{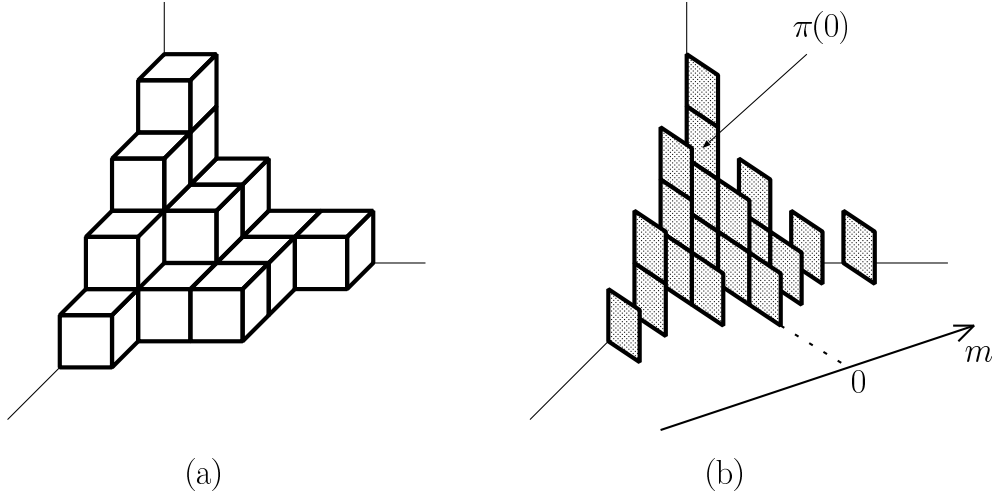}
\end{center}
\caption{Diagonal slices (b) of plane partition (a)}
\label{fig:slices}
\end{figure}

Given a plane partition $\pi = (\pi_{ij})_{i,j=1}^\infty$, 
the partition 
\beqnn
  \pi(m) = 
  \left\{\begin{array}{lcl}
    (\pi_{i,i+m})_{i=1}^\infty &\mbox{if}& m \ge 0\\
    (\pi_{j-m,j})_{j=1}^\infty &\mbox{if}& m < 0 
  \end{array}\right.
\eeqnn
is called the $m$-th diagonal slice of $\pi$. 
These partitions $\{\pi(m)\}_{m=-\infty}^\infty$ 
represent a sequence of 2D Young diagrams 
that are literally obtained by slicing 
the 3D Young diagrams (Figure\ref{fig:slices}).  

The diagonal slices are not arbitrary but satisfy 
the condition \cite{Okounkov-Reshetikhin03,ORV03} 
\beq
  \cdots \prec \pi(-2) \prec \pi(-1) \prec \pi(0) 
  \succ \pi(1) \succ \pi(2) \succ \cdots, 
\eeq
where ``$\succ$'' denotes 
{\it interlacing relation}, namely, 
\beqnn
  \lambda = (\lambda_1,\lambda_2,\ldots) 
  \succ \mu = (\mu_1,\mu_2,\ldots) 
  \;\stackrel{\mathrm{def}}{\Longleftrightarrow}\; 
  \lambda_1 \ge \mu_1 \ge \lambda_2 \ge \mu_2 \ge \cdots. 
\eeqnn
Because of these interlacing relations, 
a pair $(T,T')$ of semi-standard tableaux is obtained 
on the main diagonal slice $\lambda = \pi(0)$ 
by putting ``$m+1$'' in boxes of the skew diagram 
$\pi(\pm m)/\pi(\pm (m+1))$.  

By this mapping $\pi \mapsto (T,T')$, 
the partition function $Z_{\mathrm{3D}}$ of 
the plane partitions can be converted to 
a triple sum over the tableau $T,T'$ and 
their shape $\lambda$: 
\beq
  Z_{\mathrm{3D}} 
  = \sum_{\lambda} 
      \sum_{T,T':\mathrm{shape}\,\lambda} q^T q^{T'}, 
\label{Z3D-triplesum}
\eeq
where 
\beqnn
  q^T = \prod_{m=0}^\infty q^{(m+1/2)|\pi(-m)/\pi(-m-1)|},\\
  q^{T'} = \prod_{m=0}^\infty q^{(m+1/2)|\pi(m)/\pi(m+1)|}. 
\eeqnn
By the well known combinatorial definition of 
the Schur functions \cite{Macdonald-book}, 
the partial sum over the semi-standard tableaux 
turn out to be a special value of the Schur functions: 
\beq
  \sum_{T:\mathrm{shape}\,\lambda} q^T 
  = \sum_{T':\mathrm{shape}\,\lambda} q^{T'} 
  = s_\lambda(q^\rho),
\eeq
where 
\beqnn
  q^\rho = (q^{1/2},q^{3/2},\ldots,q^{n+1/2},\cdots). 
\eeqnn
Thus the partition function can be eventually 
rewritten as 
\beq
  Z_{\mathrm{3D}} = \sum_{\lambda} s_\lambda(q^\rho)^2. 
\eeq
Let us note that the special value of 
the Schur functions has the so called Hook formula 
\cite{Macdonald-book}
\beq
  s_\lambda(q^\rho) 
  = q^{n(\lambda)+|\lambda|/2}
    \prod_{(i,j)\in\lambda}(1 - q^{h(i,j)})^{-1}, 
\eeq
where $(i,j)$ stands for the $(i,j)$-th box 
in the Young diagram, and $n(\lambda)$ is given by 
\beqnn
  n(\lambda) = \sum_{i=1}^\infty (i-1)\lambda_i. 
\eeqnn

\subsection{Melting crystal model with external potentials}

We now deform the foregoing melting crystal model 
by introducing the external potentials
\beqnn
  \Phi_k(\lambda,p) = \sum_{i=1}^\infty q^{k(p+\lambda_i-i+1)} 
  - \sum_{i=1}^\infty q^{k(-i+1)} 
\eeqnn
with coupling constants $t_k$, $k = 1,2,3,\ldots$, 
on the main diagonal slice $\lambda = \pi(0)$. 
The right hand side of the definition of 
$\Phi_k(\lambda,p)$ is understood to be a finite sum 
(hence a rational function of $q$) by cancellation 
of terms between the two sums: 
\beqnn
  \Phi_k(\lambda,p) 
  = \sum_{i=1}^{\infty} (q^{k(p+\lambda_i-i+1)} - q^{k(p-i+1)}) 
    + q^k\frac{1-q^{pk}}{1-q^k}. 
\eeqnn
The partition function of the deformed model reads 
\beq
  Z_p(t) = \sum_{\pi} q^{|\pi|}e^{\Phi(t,\pi(0),p)}, 
\eeq
where 
\beqnn
  \Phi(t,\lambda,p) = \sum_{k=1}^\infty t_k\Phi_k(\lambda,p). 
\eeqnn
We can repeat the previous calculations in this setting 
to rewrite the new partition function $Z_p(t)$ as 
\beq
  Z_p(t) 
  = \sum_{\lambda} s_\lambda(q^\rho)^2 q^{\Phi(t,\lambda,p)}. 
\eeq

Modifying this partition function slightly, 
we obtain the instanton sum of 5D $\mathcal{N}=1$ 
SUSY $U(1)$ Yang-Mills theory \cite{MNTT04a}: 
\beq
  Z_p(t) 
  = \sum_{\pi} q^{|\pi|}Q^{\pi(0)}e^{\Phi(t,\pi(0),p)} 
  = \sum_{\lambda} s_\lambda(q^\rho)^2 
    Q^{|\lambda|}e^{\Phi(t,\lambda,p)}. 
\label{Zp-5DSYM}
\eeq
$q$ and $Q$ are related to physical parameters 
$R,\Lambda,\hbar$ of 5D Yang-Mills theory as 
\beqnn
  q = e^{-R\hbar}, \quad Q = (R\Lambda)^2. 
\eeqnn
The external potentials represent 
the contribution of Wilson loops 
along the fifth dimension \cite{BLN97}. 
In this sense, $Z_p(t)/Z_p(0)$ is a generating function 
of correlation functions of those Wilson loop operators. 

Our goal is to show that the partition function 
$Z_p(t)$ is, up to a simple factor, the tau function 
of the (1D) Toda hierarchy.  To this end, 
we now consider a fermionic representation 
of this partition function.

\section{Fermionic formula of partition function}

\subsection{Complex fermion system}

Let $\psi(z)$ and $\psi^*(z)$ denote complex 2D fermion fields 
\beqnn
  \psi(z) = \sum_{m=-\infty}^\infty \psi_mz^{-m-1}, \quad 
  \psi^*(z) = \sum_{m=-\infty}^\infty \psi^*_mz^{-m}. 
\eeqnn
The Fourier modes $\psi_m$ and $\psi^*_m$ 
of $\psi(z)$ and $\psi^*(z)$ satisfy 
the anti-commutation relations 
\beqnn
\{\psi_m,\psi^*_n\} = \delta_{m+n,0}, \quad 
\{\psi_m,\psi_n\} = \{\psi^*_m,\psi^*_n\} = 0. 
\eeqnn

The Fock space $F$ splits into charge $p$ subspaces $F_p$: 
\beqnn
  F = \bigoplus_{p=-\infty}^\infty F_p. 
\eeqnn
The charge $p$ subspace $F_p$ has a unique 
normalized ground state (charge $p$ vacuum) 
$|p\rangle$ and an orthonormal basis 
$|\lambda;p\rangle$ labeled by 
partitions $\lambda$.  $|p\rangle$ is 
characterized by the vacuum condition 
\beqnn
  \psi_m|p\rangle = 0 \quad \mbox{for}\; m \ge -p, \quad 
  \psi^*_m|p\rangle = 0 \quad \mbox{for}\; m \ge p+1. 
\eeqnn
If the partition is of the form $\lambda 
= (\lambda_1,\ldots,\lambda_n,0,0,\ldots)$, 
the associated element $|\lambda;p\rangle$ 
of the basis is obtained from $|p\rangle$ 
by the action of fermion operators as 
\beqnn
  |\lambda;p\rangle 
= \psi_{-(p+\lambda_1-1)-1}\cdots\psi_{-(p+\lambda_n-n)-1} 
    \psi^*_{(p-n)+1}\cdots\psi^*_{(p-1)+1}|p\rangle. 
\eeqnn
They are orthonormal in the sense that 
their inner products have the normalized values 
\beqnn
  \langle\lambda;p|\mu;q\rangle 
  = \delta_{pq}\delta_{\lambda\mu}. 
\eeqnn

\subsection{$U(1)$ current and fermionic representation of tau function}

The $U(1)$ current $J(z)$ of 
the complex fermion system is defined as 
\beqnn
  J(z) = {:}\psi(z)\psi^*(z){:} 
  = \sum_{k=-\infty}^\infty J_mz^{-m-1}, 
\eeqnn
where ${:}\quad{:}$ denotes the normal ordering 
with respect to the vacuum $0\rangle$: 
\beqnn
  {:}\psi_m\psi^*_n{:} 
  = \psi_m\psi^*_n - \langle 0|\psi_m\psi^*_n|0\rangle. 
\eeqnn
The Fourier modes 
\beqnn
  J_m = \sum_{n=-\infty}^\infty{:}\psi_{m-n}\psi^*_n{:}
\eeqnn
of $J(z)$ satisfy the commutation relations 
\beq
  [J_m,J_n] = m\delta_{m+n,0} 
\eeq
of the $A_\infty$ Heisenberg algebra, 
and play the role of ``Hamiltonians'' 
in the usual fermionic formula of 
the KP and 2D Toda hierarchies 
\cite{Jimbo-Miwa83,Takebe91}.  
For the case of tau functions $\tau(t,\bar{t})$, 
$t = (t_1,t_2,\ldots)$, 
$\bar{t} = (\bar{t}_1,\bar{t}_2,\ldots)$, 
of the 2D Toda hierarchy, the fermionic formula reads 
\beq
  \tau_p(t,\bar{t}) 
  = \langle p| \exp(\sum_{m=1}^\infty t_mJ_m) 
    g \exp(- \sum_{m=1}^\infty\bar{t}_mJ_{-m}) 
    |p \rangle 
\label{2DToda-tau}
\eeq
where $g$ is an element of the infinite dimensional 
Clifford group $GL(\infty)$.

\subsection{Fermionic representation of $Z_p(t)$}

The partition function $Z_p(t)$ of the deformed 
melting crystal has a fermionic representation 
of the form
\beq
  Z_p(t) = \langle p| G_{+}e^{H(t)}G_{-}|p\rangle.  
\label{Zp-fermion}
\eeq
Let us explain the constituents of this formula 
along with an outline of the derivation of 
this formula.   

$H(t)$ is the linear combination 
\beqnn
  H(t) = \sum_{k=1}^\infty t_kH_k
\eeqnn
of the ``Hamiltonians'' 
\beqnn
  H_k = \sum_{n=-\infty}^\infty q^{kn}{:}\psi_{-n}\psi^*_n{:}. 
\eeqnn
The aforementioned basis elements $|\lambda;p\rangle$ 
of the Fermion Fock space turn out to be eigenvectors 
of these Hamiltonians.  The eigenvalues are nothing 
but the the potential functions $\Phi_k(\lambda,p)$: 
\beq
  H_k |\lambda;p\rangle 
  = \Phi_k(\lambda,p) |\lambda;p\rangle. 
\eeq  

$G_{\pm}$ are $GL(\infty)$ elements of 
the special form 
\beqnn
  G_{\pm} = \exp\Bigl(\sum_{k=1}^\infty 
    \frac{q^{k/2}}{k(1-q^k)}J_{\pm k}\Bigr). 
\eeqnn
Since the numerical factors $q^{k/2}/(1-q^k)$ 
in this definition can be expanded as 
\beqnn
  \frac{q^{k/2}}{1-q^k} 
  = \sum_{m=-\infty}^{-1} q^{-k(m+1/2)} 
  = \sum_{m=0}^\infty q^{k(m+1/2)}, 
\eeqnn
one can factorize these operators as 
\beq
  G_{+} = \prod_{m=-\infty}^{-1}\Gamma_{+}(m), \quad 
  G_{-} = \prod_{m=0}^\infty \Gamma_{-}(m), 
\eeq
where 
\beqnn
  \Gamma_{\pm}(m) 
  = \exp\Bigl(\sum_{k=1}^\infty 
      \frac{1}{k}q^{\mp k(m+1/2)}J_{\pm k}\Bigr). 
\eeqnn
These $\Gamma_{\pm}(m)$'s are a specialization of 
the so called vertex operators 
\beqnn
  V_{\pm}(z) 
  = \exp\Bigl(\sum_{k=1}^\infty 
      \frac{z^k}{k}J_{\pm k}\Bigr) 
\eeqnn
for bosonization of the complex fermions. 

Following the idea of Okounkov and Reshetikhin 
\cite{Okounkov-Reshetikhin03}, 
we now consider $m$ as a fictitious ``time'' 
variable.  A plane partition then may be thought 
of as the ``path'' (or ``world volume'') of 
discrete time evolutions of a partition $\lambda$ 
that starts from the empty partition $\emptyset 
= (0,0,\ldots)$ at infinite past and ends again 
in $\emptyset$ at infinite future 
(see Figure\ref{fig:slices}).  
The vertex operators $\Gamma_{\pm}(m)$ play 
the role of transfer matrices between 
neighboring diagonal slices.  

The vertex operators $\Gamma_{\pm}(m)$ 
act on the aforementioned 
orthonormal bases $|\lambda;p\rangle$ 
and $\langle\lambda;p|$ as 
\beq
  \langle\lambda;p|\Gamma_{+}(m)
  = \sum_{\mu\succ\lambda}
    \langle\mu;p|q^{-(m+1/2)(|\mu|-|\lambda|)}
\eeq
for $m = -1,-2,\ldots$ and 
\beq
  \Gamma_{-}(m)|\lambda;p\rangle 
  = \sum_{\mu\succ\lambda} 
    q^{(m+1/2)(|\mu|-|\lambda|)}|\mu;p\rangle
\eeq
for $m = 0,1,\ldots$ \cite{Okounkov-Reshetikhin03,ORV03}.  
The right hand side of these formulas 
give a linear combination of all possible 
time evolutions of the $m$-th slice 
$\lambda = \pi(m)$ at the next time.  
The weight $q^{\mp(m+1/2)(|\mu|-|\lambda|)}$ 
of each state on the right hand side 
are exactly the factors assigned to 
the boxes of $\pi(m)/\pi(m\mp 1)$ 
in the definition of the weights $q^T,q^{T'}$ 
that appear in the combinatorial formula 
(\ref{Z3D-triplesum}). 

Since $G_{\pm}$ are products of these 
slice-to-slice ``transfer matrices'', 
$\langle p|G_{+}$ and $G_{-}|p\rangle$ 
become linear combinations of the states 
$\langle\lambda;p|$ and $|\lambda;p\rangle$ 
that evolve from the ground states $\langle p|$ 
and $|p\rangle$ at $m = \mp\infty$.  
By what we have seen above, the weights of 
$\langle\lambda;p|$ and $|\lambda;p\rangle$ 
in these linear combinations are given by 
the partial sums of $q^T$ and $q^{T'}$ over 
all semi-standard tableaux $T$ and $T'$ 
of shape $\lambda$, namely, the special value 
$s_\lambda(q^\rho)$ of the Schur function.  
Thus $\langle p|G_{+}$ and $G_{-}|p\rangle$ 
can be expressed as 
\beq
  \langle p|G_{+} 
  = \sum_{\lambda}\sum_{T:\mathrm{shape}\,\lambda}q^T 
    \langle \lambda;p| 
  = \sum_{\lambda} s_\lambda(q^\rho) \langle \lambda;p|,\\
  G_{-}|p\rangle 
  = \sum_{\lambda}\sum_{T':\mathrm{shape}\,\lambda}q^{T'} 
    |\lambda;p\rangle 
  = \sum_{\lambda} s_\lambda(q^\rho) |\lambda;p\rangle.
\eeq
The expectation value of $e^{H(t)}$ with respect 
to these states yields the fermionic representation 
(\ref{Zp-fermion}) of the partition function $Z_p(t)$.  

The fermionic representation (\ref{Zp-fermion}) 
is apparently different from the fermionic formula 
(\ref{2DToda-tau}) of tau functions of 
the 2D Toda hierarchy.  To show that $Z_p(t)$ 
is indeed a tau function, we have to rewrite 
(\ref{Zp-fermion}) to the form of (\ref{2DToda-tau}).
This is the place where the quantum torus Lie algebra 
joins the game.

\section{Quantum torus Lie algebra}

\subsection{Fermionic realization of quantum torus Lie algebra}

Let $V^{(k)}_m$ ($k = 0,1,\ldots,\; m \in \ZZ$) 
denote the following fermion bilinear forms: 
\beqnn
  V^{(k)}_m 
&=& q^{-km/2}\sum_{n=-\infty}^\infty 
  q^{kn}{:}\psi_{m-n}\psi^*_n{:}\\
&=& q^{k/2}\oint\frac{dz}{2\pi i}
  z^m{:}\psi(q^{k/2}z)\psi^*(q^{-k/2}z){:} 
\eeqnn
Note that 
\beqnn
  J_m = V^{(0)}_m, \quad H_k = V^{(k)}_0. 
\eeqnn
Actually, $V^{(k)}_m$  coincides with Okounkov 
and Pandharipande's operator $\mathcal{E}_m(z)$ 
\cite{OP02a,OP02b} specialized to $z = q^k$.  
As they found for $\mathcal{E}_m(z)$, 
our $V^{(k)}_m$'s satisfy the commutation relations
\beq
  [V^{(k)}_m,\, V^{(l)}_n] 
  = (q^{(lm-kn)/2}-q^{(kn-lm)/2})
    (V^{(k+l)}_{m+n} - \delta_{m+n,0}\frac{q^{k+l}}{1-q^{k+l}}). 
\eeq
This is a (central extension of) $q$-deformation 
of the Poisson algebra of functions on a 2-torus. 
We refer to this Lie algebra as ``quantum torus 
Lie algebra''.  More precisely, a full quantum torus 
Lie algebra should contain elements for $k<0$ as well; 
for several reasons, we shall not include those elements.

\subsection{Shift symmetry among basis of quantum torus Lie algebras}

The following relations, which we call ``shift symmetry'', 
play a central role in identifying $Z_p(t)$ as a tau function: 
\beq
G_{-}G_{+}
\Bigl(V^{(k)}_m - \delta_{m,0}\frac{q^k}{1-q^k}\Bigr)
(G_{-}G_{+})^{-1} 
= (-1)^k\Bigl(V^{(k)}_{m+k} - \delta_{m+k,0}\frac{q^k}{1-q^k}\Bigr)
\label{shift-symmetry}
\eeq
These relations are derived as follows. 

Let us recall that the fermion fields $\psi(z),\psi^*(z)$ 
transform under adjoint action by $J_{\pm k}$'s as 
\beq
  \exp\Bigl(\sum_{k=1}^\infty c_kJ_{\pm k}\Bigr)\psi(z) 
  \exp\Bigl(- \sum_{k=1}^\infty c_kJ_{\pm k}\Bigr) 
  = \exp\Bigl(\sum_{k=1}^\infty c_kz^{\pm k}\Bigr)\psi(z), \\
  \exp\Bigl(\sum_{k=1}^\infty c_kJ_{\pm k}\Bigr)\psi^*(z) 
  \exp\Bigl(- \sum_{k=1}^\infty c_kJ_{\pm k}\Bigr) 
  = \exp\Bigl(- \sum_{k=1}^\infty c_kz^{\pm k}\Bigr)\psi^*(z). 
\eeq
By letting $c_k = q^{k/2}/(1 - q^k)$, 
the exponential operators in these formulas 
turn into $G_{\pm}$, so that we have 
the operator identities 
\beq
  G_{+}\psi(z)G_{+}{}^{-1} 
&=& (q^{1/2}z;q)_\infty^{-1} \psi(z),\\ 
  G_{+}\psi^*(z)G_{+}{}^{-1} 
&=& (q^{1/2}z;q)_\infty \psi^*(z),\\
  G_{-}\psi(z)G_{-}{}^{-1} 
&=& (q^{1/2}z^{-1};q)_\infty^{-1} \psi(z), \\
  G_{-}\psi^*(z)G_{-}{}^{-1} 
&=& (q^{1/2}z^{-1};q)_\infty \psi^*(z), 
\eeq
where $(z;q)_\infty$ denotes the standard $q$-factorial symbol 
\beqnn
  (z;q)_\infty = \prod_{n=0}^\infty (1 - zq^n). 
\eeqnn

We use these operator identities to derive 
transformation of the fermion bilinear forms 
${:}\psi(q^{k/2}z)\psi^*(q^{-k/2}z){:}$ 
under conjugation by $G_{\pm}$.  Since 
\beqnn
  {:}\psi(q^{k/2}z)\psi^*(q^{-k/2}z){:} 
  = - \psi^*(q^{-k/2}z)\psi(q^{k/2}z) 
  + \frac{q^{k/2}}{(1 - q^k)z}, 
\eeqnn
let us first consider $\psi^*(q^{-k/2}z)\psi(q^{k/2}z)$. 
Under conjugation by $G_{+}$, it transforms as 
\beqnn
\lefteqn{
G_{+}\psi^*(q^{-k/2}z)\psi(q^{k/2}z)G_{+}{}^{-1} 
} \nonumber \\
&=& \frac{(q^{1/2}\cdot q^{-k/2}z;q)_\infty}
         {(q^{1/2}\cdot q^{k/2}z;q)_\infty}
    \psi^*(q^{-k/2}z)\psi(q^{k/2}z) 
\nonumber \\
&=& \prod_{m=1}^k (1 - zq^{(k+1)/2-m}) 
    \psi^*(q^{-k/2}z)\psi(q^{k/2}z). 
\eeqnn
This implies that 
\beq
\lefteqn{
G_{+}\Bigl({:}\psi(q^{k/2}z)\psi^*(q^{-k/2}z){:}
  - \frac{q^{k/2}}{(1 - q^k)z}\Bigr)G_{+}{}^{-1}
} \nonumber\\
&=& \prod_{m=1}^k (1 - q^{(k+1)/2-m}z) 
    \Bigl({:}\psi(q^{k/2}z)\psi^*(q^{-k/2}z){:} 
      - \frac{q^{k/2}}{(1 - q^k)z}\Bigr). 
\eeq
In much the same way, we can derive a similar 
transformation under conjugation by $G_{-}$. 
In this case, it is more convenient to 
rewrite the result as follows: 
\beq
\lefteqn{ 
G_{-}{}^{-1}
\Bigl({:}\psi(q^{k/2}z)\psi^*(q^{-k/2}z){:}
  - \frac{q^{k/2}}{(1 - q^k)z}\Bigr)G_{-}
} \nonumber\\
&=& \prod_{m=1}^k (1 - q^{-(k+1)/2+m}z^{-1}) 
    \Bigl({:}\psi(q^{k/2}z)\psi^*(q^{-k/2}z){:} 
      - \frac{q^{k/2}}{(1 - q^k)z}\Bigr). 
\eeq

We note here that the prefactors 
on the right hand side of the last two equations 
are related as 
\beqnn
    \prod_{m=1}^k (1 - q^{(k+1)/2-m}z) 
    = (-z)^k \prod_{m=1}^k (1 - q^{-(k+1)/2+m}z^{-1}). 
\eeqnn
Accounting for this simple, but significant 
relation, we can derive the identity 
\beq
&& 
G_{-}G_{+}
\Bigl({:}\psi(q^{k/2}z)\psi^*(q^{-k/2}z){:}
  - \frac{q^{k/2}}{(1 - q^k)z}\Bigr)
(G_{+}G_{-})^{-1} 
\nonumber \\
&=&
(-z)^k
\Bigl({:}\psi(q^{k/2}z)\psi^*(q^{-k/2}z){:}
  - \frac{q^{k/2}}{(1 - q^k)z}\Bigr). 
\eeq
The shift symmetry (\ref{shift-symmetry}) follows 
immediately from this identity. 

When $m = 0$ and $m = -k$, (\ref{shift-symmetry}) 
takes the particular form
\beq
G_{-}G_{+}\Bigl(V^{(k)}_0 - \frac{q^k}{1-q^k}\Bigr)
(G_{-}G_{+})^{-1} 
= (-1)^kV^{(k)}_k, 
\label{shift-symm(1)}\\
(G_{-}G_{+})^{-1}\Bigl(V^{(k)}_0 - \frac{q^k}{1-q^k}\Bigr)
G_{-}G_{+} 
= (-1)^kV^{(k)}_{-k}. 
\label{shift-symm(2)}
\eeq
It is these identities that we shall use 
to convert the fermionic representation of 
$Z_p(t)$ to the standard fermionic formula of 
tau functions.

\section{Integrable structure of melting crystal model}

\subsection{Partition function as tau function of 2D Toda hierarchy}

Let us split the operator $G_{+}e^{H(t)}G_{-}$ 
in (\ref{Zp-fermion}) into three pieces as 
\beqnn
  G_{+}e^{H(t)}G_{-} 
&=& G_{+}e^{H(t)/2}e^{H(t)/2}G_{-} \\
&=& G_{+}e^{H(t)/2}G_{+}{}^{-1}\cdot
    G_{+}G_{-}\cdot
    G_{-}{}^{-1}e^{H(t)/2}G_{-}
\eeqnn
and use the special cases (\ref{shift-symm(1)}) 
and (\ref{shift-symm(2)}) of the shift symmetry 
to rewrite those pieces.  

To this end, it is convenient to rewrite 
(\ref{shift-symm(1)}) and (\ref{shift-symm(2)}) as 
\beqnn
G_{+}\Bigl(H_k - \frac{q^k}{1-q^k}\Bigl)G_{+}{}^{-1} 
&=& (-1)^k G_{-}{}^{-1}V^{(k)}_kG_{-},\\
G_{-}{}^{-1}\Bigl(H_k - \frac{q^k}{1-q^k}\Bigr)G_{-} 
&=& (-1)^k G_{+}V^{(k)}_{-k}G_{+}{}^{-1}.
\eeqnn
Though the operators $V^{(k)}_{\pm k}$ on 
the right hand side are unfamiliar in the theory 
of integrable hierarchies, we can convert them 
to the familiar ``Hamiltonians'' $J_{\pm k} 
= V^{(0)}_{\pm k}$ of the Toda hierarchies as 
\beq
  q^{W/2}V^{(k)}_k q^{-W/2} = V^{(0)}_k = J_k,\quad 
  q^{-W/2}V^{(k)}_{-k} q^{W/2} = V^{(0)}_{-k} = J_{-k}, 
\label{WJ-relation}
\eeq
where $W$ is a special element of $W_\infty$ algebra: 
\beqnn
  W = W^{(3)}_0 
  = \sum_{n=-\infty}^\infty n^2{:}\psi_{-n}\psi^*_n{:}
\eeqnn
We thus eventually obtain the relations 
\beq
    G_{+}\Bigl(H_k - \frac{q^k}{1-q^k}\Bigl)G_{+}{}^{-1} 
&=& (-1)^k G_{-}{}^{-1}q^{-W/2}J_k q^{W/2}G_{-},\\
    G_{-}{}^{-1}\Bigl(H_k - \frac{q^k}{1-q^k}\Bigr)G_{-} 
&=& (-1)^k G_{+}q^{W/2}J_{-k}q^{-W/2}G_{+}{}^{-1} 
\eeq
between $H_k$'s and $J_{\pm k}$'s.  

By these relations, $G_{+}e^{H(t)/2}G_{+}{}^{-1}$  
can be calculated as 
\beqnn
&&G_{+}e^{H(t)/2}G_{+}{}^{-1}\\
&=& \exp\Bigl(\sum_{k=1}^\infty \frac{t_kq^k}{2(1-q^k)}\Bigr)
   G_{-}{}^{-1}q^{-W/2}
   \exp\Bigl(\sum_{k=1}^\infty \frac{(-1)^kt_k}{2}J_k\Bigr)
   q^{W/2}G_{-}. 
\eeqnn
A similar expression can be derived for 
$G_{-}{}^{-1}e^{H(t)}G_{-}$ as well.  
We can thus rewrite $G_{+}e^{H(t)}G_{-}$ as 
\beqnn
  G_{+}e^{H(t)}G_{-} 
= \exp\Bigl(\sum_{k=1}^\infty\frac{t_kq^k}{1-q^k}\Bigr) 
  G_{-}{}^{-1}q^{-W/2} 
  \exp\Bigl(\sum_{k=1}^\infty\frac{(-1)^kt_k}{2}J_k\Bigr) 
\times {} \nonumber \\
\mbox{} \times g 
  \exp\Bigl(\sum_{k=1}^\infty\frac{(-1)^kt_k}{2}J_{-k}\Bigr) 
  q^{-W/2}G_{+}{}^{-1} 
\eeqnn
where 
\beq
   g = q^{W/2}(G_{-}G_{+})^2 q^{W/2}. 
\eeq

The partition function $Z_p(t)$ is given by the expectation 
value of this operator with respect to $\langle p|$ 
and $|p\rangle$.  Since the action by the leftmost 
and rightmost pieces of $g$ yields only a scalar 
multiplier to $\langle p|$, $|p\rangle$ as 
\beq
  \langle p|G_{-}{}^{-1}q^{-W/2} 
&=& q^{-p(p+1)(2p+1)/12} \langle p|, \\
  q^{-W/2}G_{+}{}^{-1}|p \rangle 
&=& q^{-p(p+1)(2p+1)/12}|p \rangle, 
\eeq
$Z_p(t)$ can be expressed as 
\beq
  Z_p(t) 
&=& \exp\Bigl(\sum_{k=1}^\infty\frac{t_kq^k}{1-q^k}\Bigr) 
  q^{-p(p+1)(2p+1)/6} \times {}
\nonumber \\
&&  \mbox{} \times 
  \langle p|\exp\Bigl(\sum_{k=1}^\infty\frac{(-1)^kt_k}{2}J_k\Bigr) 
  g \exp\Bigr(\sum_{k=1}^\infty\frac{(-1)^kt_k}{2}J_{-k}\Bigr) 
  |p \rangle. 
\label{Zp-tau}
\eeq
The expectation value $\langle p|\cdots|p \rangle$ 
takes exactly the form of (\ref{2DToda-tau}).  
Thus, up to the simple prefactor, $Z_p(t)$ is essentially 
a tau function of the 2D Toda hierarchy.  
Thus we find that an integrable structure behind 
the melting crystal model is the 2D Toda hierarchy.  
This is, however, not the end of the story.

\subsection{1D Toda hierarchy as true integrable structure}

The foregoing calculation is based on the splitting 
\beqnn
  G_{+}e^{H(t)}G_{-} 
  = G_{+}e^{H(t)/2}G_{+}{}^{-1}\cdot
    G_{+}G_{-}\cdot
    G_{-}{}^{-1}e^{H(t)/2}G_{-}. 
\eeqnn
Actually, we could have started from a different splitting 
of $G_{+}e^{H(t)}G_{-}$, e.g., 
\beqnn
  G_{+}e^{H(t)}G_{-} 
  = G_{+}e^{H(t)}G_{+}{}^{-1}\cdot G_{+}G_{-} 
  = G_{+}G_{-}\cdot G_{-}{}^{-1}e^{H(t)}G_{-}
\eeqnn
This leads to another set of expressions of $Z_p(t)$ 
in which only the $\langle p|\cdots|p \rangle$ part 
is different.  We thus have the following three 
different expressions for this part: 
\beq
\lefteqn{
  \langle p|\exp\Bigl(\sum_{k=1}^\infty\frac{(-1)^kt_k}{2}J_k\Bigr) 
  g \exp\Bigr(\sum_{k=1}^\infty\frac{(-1)^kt_k}{2}J_{-k}\Bigr) 
  |p \rangle 
} \nonumber \\
&=&  \langle p|
  \exp\Bigl(\sum_{k=1}^\infty (-1)^kt_kJ_k\Bigr) g 
  |p \rangle 
\nonumber \\
&=& \langle p|
  g \exp\Bigl(\sum_{k=1}^\infty (-1)^kt_kJ_{-k}\Bigr) 
  |p \rangle. 
\eeq

These identities of the expectation values can be 
directly derived from the operator identities 
\beq
  J_kg = gJ_{-k}, \; k = 1,2,3,\ldots 
\label{Jg=gJ}
\eeq
satisfied by $g$.  (These operator identities 
themselves are a consequences of the shift symmetry 
of $V^{(k)}_m$'s.)  Generally speaking, this kind of 
operator identities imply symmetry constrains 
on the tau functions \cite{NTT95,Takasaki96}; 
in the present case, the constraints read 
\beq
  \frac{\rd}{\rd t_k}\tau_p(t,\bar{t}) 
  + \frac{\rd}{\rd\bar{t}_k}\tau_p(t,\bar{t}) 
  = 0, \quad 
  k = 1,2,3,\ldots. 
\eeq
In other words, the tau function is a function 
of $t - \bar{t}$, 
\beqnn
  \tau_p(t,\bar{t}) 
= \tau_p(t-\bar{t},0) 
= \tau_p(0,\bar{t}-t), 
\eeqnn
and reduces to a tau function $\tau_p(t)$ of 
the 1D Toda hierarchy that has a single series 
of time variables $t = (t_1,t_2,\ldots)$ 
rather than the two series of the 2D Toda hierarchy.  
Thus the 1D Toda hierarchy eventually turns out 
to be an underlying integrable structure 
of the deformed melting crystal model.  

The same conclusion can be derived for 
the instanton sum (\ref{Zp-5DSYM}) of 
5D SUSY $U(1)$ Yang-Mills theory.  
It has a fermionic representation of the form 
\beq
  Z_p(t) 
  = \langle p|G_{+}Q^{L_0}e^{H(t)}G_{-}|p \rangle 
\label{Zp-5DSYM-fermion}
\eeq
where $L_0$ is a special element of the Virasoro algebra: 
\beqnn
  L_0 
  = \sum_{n=-\infty}^\infty n{:}\psi_{-n}\psi^*_n{:}. 
\eeqnn
One can repeat almost the same calculations 
as the previous case to convert $Z_p(t)$ to 
the form of (\ref{Zp-tau}). 
The counterpart of $g$ is given by 
\beq
  g = q^{W/2}G_{-}G_{+}Q^{L_0}G_{-}G_{+}q^{W/2}, 
\eeq
which, too, satisfy the reduction conditions 
(\ref{Jg=gJ}) to the 1D Toda hierarchy.  
Thus a relevant integrable structure 
is again the 1D Toda hierarchy.

\section{Concluding remarks}

\subsection{Problems on 4D instanton sum} 

In deriving the instanton sum (\ref{Zp-5DSYM}), 
5D space-time is partially compactified 
in the fifth dimension as $\RR^4 \times S^1$. 
The parameter $R$ is the radius of $S^1$. 
Therefore, letting $R \to 0$ amounts to 4D limit.  

Unfortunately, it is not straightforward 
to achieve such a 4D limit in the present setup.  
Firstly, the 5D instanton sum with external potentials 
does not have a reasonable limit as $R \to 0$. 
Any naive prescription letting $R \to 0$ yields 
a result in which $t$ dependence disappears 
or becomes trivial \cite{Nakatsu-Takasaki07}.  
Secondly, the shift symmetry of the quantum torus 
Lie algebra ceases to exist in the limit as 
$q = e^{-R\hbar} \to 1$.  Speaking more precisely, 
the quantum torus Lie algebra itself turns into 
a $W_\infty$ algebra in this limit, but no analogue 
of shift symmetry (\ref{shift-symmetry}) is known 
for the latter case.  For these reasons, 
the 4D case has to be studied independently. 

The 4D instanton sum 
\cite{Nekrasov02,Nakajima-Yoshioka05}, 
too, is a sum over partitions.  Moreover, 
this statistical sum has a fermionic representation 
\cite{Losev-etal03,Nekrasov-Okounkov03}.  
Marshakov and Nekrasov 
\cite{Marshakov-Nekrasov06,Marshakov07} 
further introduced external potentials therein.  
Actually, the 4D instanton sum for $U(1)$ gauge theory 
is almost identical to the generating function 
of Gromov-Witten invariants of $\CC\PP^1$ 
\cite{OP02a,OP02b}.  This can be most clearly seen 
in the fermionic representation of 
these generating functions, which reads 
\beq
  Z^{\mathrm{4D}}_p(t) 
  = \langle p| e^{J_1/\hbar}
    \exp\Bigl(\sum_{k=1}^\infty 
      t_k\frac{\mathcal{P}_{k+1}}{k+1}\Bigr) 
    e^{J_{-1}/\hbar} |p\rangle, 
\eeq
where $\mathcal{P}_k$'s are fermion bilinear forms  
introduced by Okounkov and Pandharipande 
for a fermionic representation of (absolute) 
Gromov-Witten invariants of $\CC\PP^1$ \cite{OP02a}.  
As regards these Gromov-Witten invariants 
(in other words, correlation functions of 
the topological $\sigma$ model) , 
it has been known for years 
\cite{Eguchi-Hori-Yang95,Eguchi-Hori-Xiong97,
Getzler00,Pandharipande00,Givental01} 
that a relevant integrable structure 
is the 1D Toda hierarchy.  

Thus the 1D Toda hierarchy is expected 
to be the integrable structure of 
the 4D instanton sum as well. 
This has been confirmed by Marshakov and Nekrasov 
in detail \cite{Marshakov-Nekrasov06,Marshakov07}. 
What is still missing, however, is a formula like 
(\ref{Zp-tau}) that directly connects $Z^{\mathrm{4D}}_p(t)$ 
with the standard fermionic formula (\ref{2DToda-tau}) 
of the tau function.  Finding a 4D analogue of 
(\ref{Zp-tau}) is thus an intriguing open problem. 
This issue is also closely related to the fate 
of shift symmetry (\ref{shift-symmetry}) 
in the $q \to 1$ limit.

\subsection{Relation to topological strings}

Our results are directly or indirectly connected 
with some aspects of topological strings as well. 

1.  According to the theory of topological vertex 
\cite{AKMV03}, the partition function $Z_p(t)$ 
of the deformed melting crystal model 
has another interpretation as the $A$-model 
topological string amplitude 
for the toric Calabi-Yau threefold 
$\mathcal{O}\oplus\mathcal{O}(-2) \to \bf{CP}^1$.  
In this interpretation, $q$ and $Q$ are parametrized 
by the string coupling constant $g_{\mathrm{st}}$ and 
the K\"ahler volume $a$ of $\CC\PP^1$ as 
\beqnn
  q = e^{-g_{\mathrm{st}}}, \quad Q = e^{-a}. 
\eeqnn
Specializing the value of $t$ leads to 
several interesting observations 
\cite{Nakatsu-Takasaki07}. 

2. A generating function of the two-legged  
topological vertex $W_{\lambda\mu} \sim 
c_{\lambda\mu\bullet}$ is known to give a 
tau function of the 2D Toda hierarchy \cite{Zhou03}.  
In the fermionic representation (\ref{2DToda-tau}), 
this amounts to the case where 
\beq
  g = q^{W/2}G_{+}G_{-}q^{W/2}.
\eeq
(Actually, for complete agreement with 
the usual convention, we have to 
replace $W$ with 
\beqnn
  K = \sum_{n=-\infty}^\infty 
      \Bigl(n - \frac{1}{2}\Bigr)^2 
      {:}\psi_{-n}\psi^*_n{:}, 
\eeqnn
but this is not a serious problem. 
The difference can be absorbed by 
rescaling $t_k$'s.)  Let us stress that 
this $GL(\infty)$ element does not satisfy 
the reduction condition (\ref{Jg=gJ}) 
to the 1D Toda hierarchy.  

3. A generating function of double Hurwitz numbers 
for coverings of $\CC\PP^1$ gives yet another type 
of tau function of the 2D Toda hierarchy 
\cite{Okounkov00}.  Actually, the $GL(\infty)$ 
element for the fermionic representation 
is given by 
\beq
  g = q^{W/2}. 
\eeq
(More precisely, as in the previous case, 
$W$ has to be replaced by $K$, but 
the difference is again irrelevant.) 
In this case, the reduction condition 
(\ref{Jg=gJ}) to the 1D Toda hierarchy 
is not satisfied, but the operator identities 
(\ref{WJ-relation}) imply that another set 
of reduction conditions are hidden behind 
(see below).  

\subsection{Constraints and quantum torus Lie algebra}

As a consequence of the shift symmetry of $V^{(k)}_m$'s, 
the $GL(\infty)$ elements $g$ of the aforementioned models 
of topological strings turn out to satisfy 
some algebraic relations other than (\ref{Jg=gJ}).  
According to general results on constraints 
of the 2D Toda hierarchy \cite{NTT95,Takasaki96},  
such relations imply the existence of 
constraints on the tau functions and 
the Lax and Orlov-Schulman operators. 
Those constraints inherit the structure 
of the quantum torus Lie algebra.  
Let us illustrate this observation 
for the case of double Hurwitz numbers 
over $\CC\PP^1$. 

The $GL(\infty)$ element $g = q^{W/2}$ 
for this case satisfies the operator identities 
\beq
  J_kg = gV^{(k)}_k, \quad 
  gJ_{-k} = V^{(k)}_{-k}g 
\eeq
as a consequence of (\ref{WJ-relation}).  
These identities can be converted 
to the constraints 
\beq
  L = q^{1/2}q^{\bar{M}}\bar{L}, \quad 
  \bar{L}^{-1} = q^{-1/2}q^ML^{-1} 
\label{Lq^M-constraint}
\eeq
on the Lax and Orlov-Schulman operators 
$L,M,\bar{L},\bar{M}$ of the 2D Toda hierarchy. 
Emergence of the exponential operators 
$q^M$ and $q^{\bar{M}}$ is a manifestation 
of the quantum torus Lie algebra.  
To see this, let us recall that the Lax and 
Orlov-Schulman operators satisfy the (twisted) 
canonical commutation relations 
\beq
  [L,M] = L, \quad [\bar{L},\bar{M}] = \bar{L}. 
\eeq
This implies that the monomials $q^{-km/2}L^{m}q^{kM}$ 
and $q^{-km/2}\bar{L}^mq^{k\bar{M}}$ 
of $L,q^M,\bar{L},q^{\bar{M}}$ give two copies 
of realizations of the quantum torus Lie algebra.  

The constraints (\ref{Lq^M-constraint}) are 
remarkably similar to the ``string equations'' 
\beq
  L = \bar{M}\bar{L}, \quad 
  \bar{L}^{-1} = ML^{-1} 
\eeq
of $c = 1$ strings at self-dual radius 
\cite{Eguchi-Kanno94,Takasaki95,NTT95,Takasaki96}. 
A relevant algebraic structure of 
these string equations is the $W_\infty$ 
algebra.  Thus (\ref{Lq^M-constraint}) 
may be thought of as $q$-deformations of 
these $W$-algebraic constraints.

\subsection*{Acknowledgements}
We are grateful to Nikita Nekrasov and Motohico Mulase 
for valuable comments and fruitful discussion.  
K.T is partly supported by Grant-in-Aid for 
Scientific Research No. 18340061 and No. 19540179 from 
the Japan Society for the Promotion of Science.

\end{document}